\documentclass[a4paper,11pt]{article}
\usepackage{fullpage}

\usepackage{cite}
\usepackage{amsmath,amssymb,amsfonts}
\usepackage{graphicx}
\usepackage{textcomp}
\usepackage{xcolor}
\usepackage{svg,import,multicol}

\usepackage[utf8]{inputenc}
\usepackage{hyperref}
\usepackage{tikz,rotating,color}
\usetikzlibrary{matrix,arrows,automata,shapes,patterns}
\usepackage{amsthm}
\newtheorem{proposition}{Observation}
\newtheorem{remark}{Remark}

\newtheorem{theorem}{Theorem}
\newtheorem{corollary}{Corollary}
\newtheorem{definition}{Definition}

\newcommand{\toto}{xxx}
\newenvironment{proofT}{\noindent{\bf
		Proof }} {\hspace*{\fill}$\Box_{Theorem~\ref{\toto}}$\par\vspace{3mm}}

\newenvironment{proofC}{\noindent{\bf
		Proof }} {\hspace*{\fill}$\Box_{Corollary~\ref{\toto}}$\par\vspace{3mm}}

\usepackage[normalem]{ulem}
\newcommand\redout{\bgroup\markoverwith
	{\textcolor{red}{\rule[0.5ex]{2pt}{0.8pt}}}\ULon}

\usepackage{algorithm}
\usepackage{rounddiag}

\usepackage{technote}
\usepackage{homodel}

\newconstruct{\FOREACH}{\textbf{for each}}{\textbf{do}}{\ENDFOREACH}{}

\newident{noop}
\newconstruct{\UPON}{\textbf{upon}}{\textbf{do}}{\ENDUPON}{}

\newcommand\Figure{\text{Fig.}}

\newcommand\selection{\ensuremath{\textsf{selection}}}
\newcommand\reward{\ensuremath{\textsf{reward}}}
\newcommand\info{\ensuremath{M}}
\newcommand\Blocks{\ensuremath{\mathbb{B}}}
\newcommand\bc{\ensuremath{\textit{bc}}}

\newconstruct{\FUNCTION}{\textbf{Function}}{\textbf{:}}{\ENDFUNCTION}{}

\title{
On Fairness in Committee-based Blockchains}

\author{Yackolley Amoussou-Guenou$^{\ddagger,\star}$, Antonella Del Pozzo$^{\ddagger}$, \\Maria Potop-Butucaru$^\star$, Sara Tucci-Piergiovanni$^\ddagger$\\~\\
	$^\ddagger$CEA LIST, PC 174, Gif-sur-Yvette, 91191, France\\
	$^\star$Sorbonne Université, CNRS, LIP6, F-75005 Paris, France \\
}
\date{}

\def\BibTeX{{\rm B\kern-.05em{\sc i\kern-.025em b}\kern-.08em
    T\kern-.1667em\lower.7ex\hbox{E}\kern-.125emX}}

\begin{document}
\newcounter{linecounter}
\newcommand{\linenumbering}{(\arabic{linecounter})}
\renewcommand{\line}[1]{\refstepcounter{linecounter}
	\label{#1}
	\linenumbering}
\newcommand{\resetline}{\setcounter{linecounter}{0}}

\maketitle
	
\begin{abstract}

Committee-based blockchains are among the most popular alternatives of \emph{proof-of-work} based blockchains, such as Bitcoin.
They provide strong consistency (no fork) under classical assumptions, and avoid using energy-consuming mechanisms to add new blocks in the blockchain.
For each block, these blockchains use a committee that executes Byzantine-fault tolerant distributed consensus
to decide the next block they will add in the blockchain.
Unlike Bitcoin, where there is only one creator per
block with high probability, in committee-based blockchain any block is
cooperatively
created. In order to incentivize committee members to participate to the
creation of new blocks
rewarding schemes have to be designed. In this paper, we study the
fairness of rewarding in
committee-based blockchains and we provide necessary and sufficient
conditions on the system communication under which it is possible to
have a fair reward mechanism.

\end{abstract}



\section{Introduction}\label{sec:intro}

Blockchain technology is one of the most appealing technology since its introduction in the Bitcoin White Paper \cite{bitcoinNakamoto} in 2008. 
A blockchain is a distributed ledger, where information are stocked in blocks and hashes link blocks in order to have a chain structure.
Blockchains systems mostly use proof-of-work, where the first process that solves a crypto-puzzle can add a new block to the blockchain.
First, this technique is highly energy consuming and second, it does not ensure consistency, \emph{i.e.} conserving the chain structure. Forks may happen and lead to a tree structure.
Some alternatives arisen to avoid at least one of these issues. 
For example, proof-of-stake based blockchains, where the probability of being able to add a block depends on the stake of a process; 
this alternative solves the energy consumption issue, but not the consistency; they are also subject to the \emph{nothing-at-stake} problem, 
where processes try to produce blocks on all the forks, to ensure some reward, and by doing so, do not resolve the forks.
However, in \cite{Saleh19}, Saleh shows that the nothing-at-stake problem is mitigated.
Other alternatives tackle both issues, for example, committee-based blockhains.
In committee-based blockchains, for each height/block, a committee is selected and that committee uses a consensus algorithm to decide on the next block to append in the blokchain. By construction, committee-based blockchains are not subject to nothing-at-stake, since the ensure consistency (no fork).

To motivate processes to add and maintain the blockchain, rewarding mechanisms are in place. 
Because committee members can be faulty, rewarding mechanisms are inherently more complex to handle and their properties  must be studied.
Ad minimum the rewarding mechanism
must be fair, \emph{i.e.}  distributing the rewards in proportion to the merit of participants,
where merit abstracts the notion of effort processes take for the construction of the blockchain \cite{ADLPT19}, for instance it models the hashing power in Bitcoin.

	Informally, we say that a blockchain protocol is fair if any \emph{correct process} (a process that followed the protocol throughout the whole execution) that has $\alpha$ fraction of the total merit in the system will get at least $\alpha$ fraction of the total reward that is given in the system. 
Our fairness analysis, in line with Francez's definition of fairness \cite{francez86},  generally defines the fairness of protocols based on voting committees (e.g. Byzcoin\cite{BizCoin}, Hyperledger \cite{hyperledger}, PeerCensus\cite{PeerCensus}, RedBelly \cite{redbelly17}, SBFT \cite{sbft18}, etc.), actually studies fairness by separating the fairness of their 
\emph{selection mechanism}  and the fairness of their \emph{reward mechanism}. 

The selection mechanism is in charge of selecting the subset of processes that will participate to the agreement on the next block to append in the blockchain, while
the reward mechanism defines the way rewards are distributed among  processes that participate to the agreement. 
We propose a formal definition of fairness of selection mechanisms, and then we study the fairness of some selection mechanisms.
The analysis of the reward mechanism  allowed to establish the following fundamental result and necessary conditions with respect to the fairness of  committee-based blockchains  as follows:

\begin{centering}
\textit{There exists a(n) (eventual) fair reward mechanism for  committee-based blockchains if and only if the system is (eventual) synchronous and processes are detectable.} (Theorems \ref{t:strongFairness} and \ref{t:eventualFairness}).
 \end{centering}
 
 The rest of the paper is as follows. 
In Section \ref{sec:relatedwork}, we compare the existing studies on fairness in blockchain systems;
in Section \ref{sec:model}, we define the system model;
in Section \ref{sec:algo}, we give a generic idea of committee-based blockchain systems, and the  consensus problems;
in Section \ref{sec:fairness}, we provide a formal definition of fairness in committee-based systems, and some analysis;
in Section \ref{sec:simulation}, we analyze some behaviors and the communication model impact on rewards; 
and in Section \ref{sec:conclusion}, we conclude.

\section{Related Work}\label{sec:relatedwork}

The closest work in blockchain systems to our fairness study (however very different in its scope) is the study of the \emph{chain-quality}.
In \cite{GarayKL15}, Garay \emph{et al.} define the notion of \emph{chain-quality} as  the proportion of blocks mined by honest miners in any given window; Garay \emph{et al.} study the conditions under which during a given window of time, there is a bounded ratio of blocks in the chain that malicious players produced, over the total blocks in the blockchain.
Kiayias \emph{et al.} in \cite{KRDO17} proposes Ourobouros\cite{KRDO17} and analyses the chain-quality property.
	In \cite{PS17}, Pass and Shi propose a notion of \emph{fairness} which is an extension of the chain-quality property, they address one of the vulnerabilities of Bitcoin studied formally in \cite{ES14, ES18}. In \cite{ES14, ES18}, Eyal and Sirer prove that if the adversary controls a coalition of miners holding even a  minority fraction of the computational power, this coalition can gain twice its share. Fruitchain \cite{PS17} overcomes this problem by ensuring that no coalition controlling
less than a majority of the computing power can gain more than a factor $1+3\delta$ by not respecting  the protocol, where $\delta$ is a parameter of the protocol.
	We note that in their model, only one process creates a block in the blockchain, and that process have a reward for the created block. 
In \cite{GW18}, Guerraoui and Wang study the effect of the delays of messages propagation in Bitcoin, and show that in a system of two miners, a miner can take advantage of the delays and be rewarded exponentially more that its expectation.
We extend the definition of \cite{PS17} for systems where each block is produced  by a subset of processes. This is the case of  Tendermint \cite{opodis18, netys19} or Hyperledger \cite{hyperledger} for example, where for each block there is a subset of processes, \emph{a committee} that produces that block. 

In \cite{GDT17, GRT18}, G{\"u}rcan \emph{et al.} study the fairness from the point of view of the processes that do not participate to the construction of the blockchain.
Herlihy and Moir do a similar work in \cite{HM16} where the authors study users' fairness and consider as an example Tendermint. 
Herlihy and Moir discussed how processes with malicious behavior could violate fairness by choosing transactions, and then they propose
modifications to the original Tendermint to make those violations detectable and accountable. In \cite{LSKM19}, Lev{-}Ari \emph{et al.} study fairness on transactions in committee-based blockchains with synchronous assumptions by using a detectable communication abstraction allowing them to identify malicious participants. Our work does not study fairness in the point of view of users.


Recent works consider the distributions of rewards in proof-of-stake based blockchains.
In particular, in \cite{FKORVW18}, Fanti \emph{et al.} define equitability, which represents the evolution of the fraction of total stake of nodes, in particular, they compute the effect of so called compounding where reward are directly re-invest in stakes. 
Karakostas \emph{et al.} define in \cite{KKNZ19} egalitarianism. Egalitarianism means that nodes, no matter their stake fraction, win the election process to append a new block the same amount. 
In this work, we focus more on fairness, where nodes that have a higher merit should get more rewards.
Our notion of fairness is different that egalitarianism, since egalitarianism's goal is to have all nodes rewarded the same, no matter their merit.

\section{System Model}\label{sec:model}

The system is composed of an infinite set $\Pi$ of sequential processes, namely $\Pi = \{p_1, p_2, \dots p_i, \dots \}$; 
$i$ is the \textit{index} of $p_i$. 
\textit{Sequential} means that a process executes one step at a time.
This does not prevent it from executing several threads with an appropriate multiplexing. As local processing time are negligible with respect to message transfer delays, we consider it as being equal to zero.

\textbf{Arrival model.}
We assume  a \textit{finite arrival model} \cite{aguilera2004}, i.e.  the system has  infinitely many processes but each run has only finitely many.   The size of the set $\Pi_{\rho} \subset \Pi$  of processes that participate  in each system run is not a priori-known.
We also consider a finite subset $V \subseteq \Pi_\rho$  of committee-members. The set $V$ may change during any system run and its size $|V|=n$ is a priori known. A process is promoted in $V$ by a selection function.  Such selection function can be based for instance on the stake in proof-of-stake blockchains, or computing power in proof-of-work blockchains. 

\textbf{Time assumptions on communication.} The processes communicate by exchanging messages through an eventually synchronous  network \cite{DLS88}. \textit{Eventually Synchronous} means that after a finite unknown time  $\tau$ there is an a priori unknown bound $\delta$ on the message transfer delay. 
When  $\tau=0$ and $\delta$ is known the network is \emph{synchronous}.

\textbf{Failure model.} Some processes can exhibit a Byzantine behavior \cite{RA80} in the system; we do not assume any bound on their number, but up to $f$ committee members can exhibit a Byzantine behavior at each point of the execution. A Byzantine process is a process that behaves arbitrarily. 
A process  that exhibits a Byzantine behavior is called a Byzantine or a \textit{faulty} process. 

Let $T$ be a period during the execution. We say that a process $p_i$ is $T$-correct, iff during the period $T$, $p_i$ follows the protocol.
A process $p_i$ is correct if $\forall T$, $p_i$ is $T$-correct.

\textbf{Communication primitives.} 
In the following, we assume the presence of a broadcast primitive.
The primitive $\textsf{broadcast} ()$ is a best effort broadcast, which means that when a correct process broadcasts a value, eventually all the correct processes deliver it.
A process $p_i$ receives a broadcast of a message by executing the primitive $\textsf{delivery}()$.
Messages are created with a digital signature, and we assume that digital signatures are unforgeable. When a process $p_i$ delivers a message, it knows the process $p_j$ that created the message.

\section{Committee-based Blockchains}\label{sec:algo}
%

Any committee-based blockchain uses instances of consensus, solving a form of repeated consensus. 
This way, each committee agrees on a single value to avoid forks. Anceaume \emph{et al.} \cite{ADLPT19} proved that consensus is required to avoid forks.

Each correct process outputs an infinite sequence of decisions called the \emph{output} of the process.
More formally, as described by Delporte-Gallet \emph{et al.} \cite{DDFPT08}, and generalized to Byzantine failures in \cite{opodis18},
%
	an algorithm implements a repeated consensus if and only if it satisfies the following properties:
	(i) \emph{Termination.} Every correct process has an infinite output.
	(ii) \emph{Agreement.} If the $i^{th}$ value of the output of a correct process is $B$, and the  $i^{th}$ value of the output of another correct process is $B'$, then $B=B'$.
	(iii) \emph{Validity.} Each value in the output of any correct process is valid with respect to a predefined predicate.

\begin{figure*}[t!]
	\centering
	\begin{tikzpicture}[thick,scale=0.8, every node/.style={scale=0.9}]
		\tikzstyle{every state}=[draw, circle, minimum size=2.6cm,node distance=.25\textwidth]
		\tikzset{initial text={\texttt{init:}$h:=1$}}
		\node[initial above,state] (compCommittee) [align=center]{Compute \\Committee \\of $h$};
		
		\node[state, draw=none, right of = compCommittee] (inv1) {};
		\node[state, above right of = inv1] (consensus) [align=center] {- Solve \\Consensus \\- Send decision};
		\node[state, below right of = inv1] (waiter) [align=center] {Wait for \\decision from \\committee};
		
		\node[state, draw=none, below right of = consensus] (inv2) {};
		\node[state, right of = inv2] (collect) [align=center] {- Wait $\Delta$ \\to collect \\more decision\\- Update $\Delta$};
		
		\draw [->] (compCommittee) edge[bend left] node[above, rotate=20, align=center]{if the process is a \\committee member} (consensus);
		
		\draw [->] (compCommittee) edge[bend right] node[below, rotate=340,align=center]{if the process is not a \\committee member} (waiter);
		
		\draw [->] (consensus) edge[bend left] node[below, rotate=340] {} (collect);
		
		\draw [->] (waiter) edge[bend right] node[below, rotate=20, align=center] {If enough evidence \\of a decision} (collect);
		
		\draw [->] (waiter) edge[loop below, looseness=4] node[below, align=center]{Not enough evidence \\of a decision} (waiter);
		
		\draw [->] (collect) edge node[above]{$h++$} (compCommittee);
	\end{tikzpicture}
	\caption{State Machine of a Repeated Consensus Algorithm to Build a Committee-based Blockchain.}
	\label{fig:repConsensus}
\end{figure*}
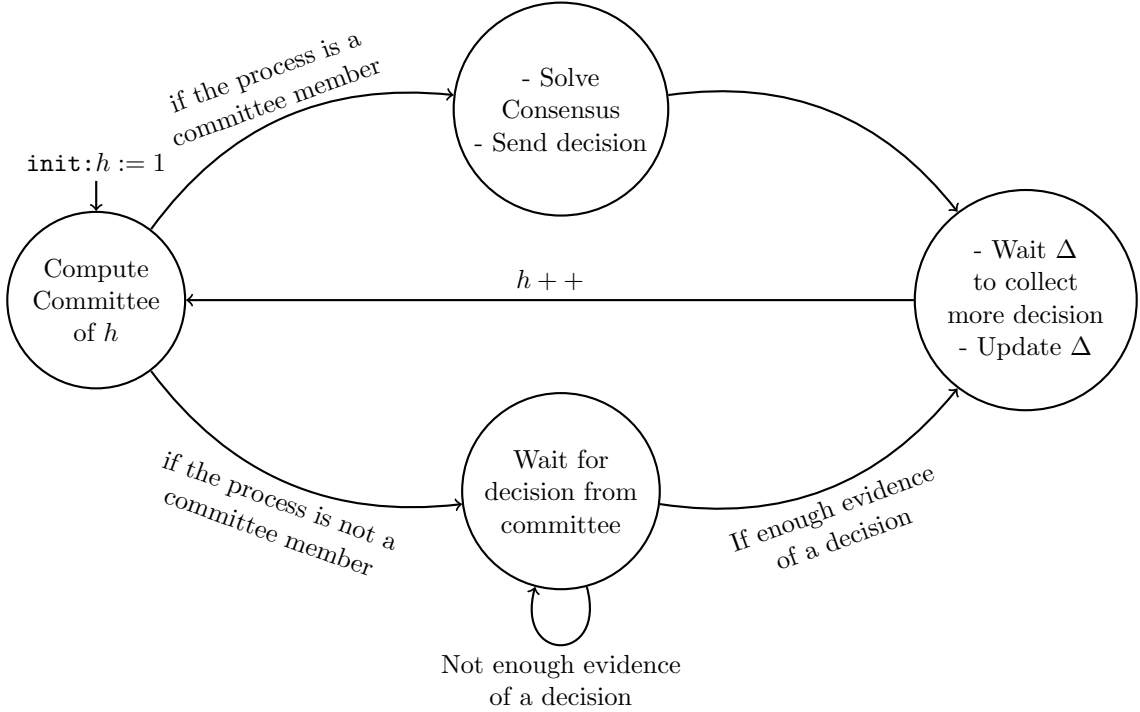

%
%
%
%
\subsection*{Detailed Description of the Algorithm}
We denote by $\Blocks$ the set of all blocks. A block contains, among other things, a header, and a list of transactions.
Let $\bc \in \Blocks^{*}$, be a finite sequence of blocks. $|\bc|$ is the length (the number of blocks) of $\bc$.
We say that $\bc$ is a \emph{blockhain} if $\forall k \in \mathbb{N}:  0 < k \le |\bc|$, in the header of the block at position $k$ in $\bc$, there is the hash of the block at position $k-1$.
If additionally, the list of transactions in each blocks in the blockchain is valid with respect to the given application, 
we say that $\bc$ is a \emph{valid} blockchain. 
The block at position $0$ is the \emph{genesis block}.
Each process has a non-negative \emph{stake}, which is the total amount of currency it has.

$\forall h >0$, let $V_h$ be the set of committee members for the height $h$. $\forall h>0$, we assume that $|V_h|=n$, the size of committee member is fixed and equal to $n$.

The state machine of the algorithm is depicted in $\Figure$ \ref{fig:repConsensus}.
The genesis block initializes the blockchain, selects the committee that will produce the block at position $1$, describes how rewards will be distributed among committee members (which we call the \emph{reward mechanism}),
gives the initial distribution of stakes, and describes how processes will be selected for being part of committees with respect to the state of the blockchain (the \emph{selection mechanism}). 
These information should be public, and known by all processes such that with the history of the blockchain, 
all processes can always compute deterministically the sets of committee members.
For a height, processes not in the corresponding committee just wait for the decision from the committee members.
The committee members for that height execute the consensus algorithm to decide upon the block for that height.
Once a process decides on a block, it sends the decided block to the whole network, and moves to the next height. 
Non-committee members wait to collect enough times the same decided block from the committee members, 
in order to be tolerant to failures, and then move to the next height.
When moving to the next height, processes wait a certain amount of time to collect more messages from committee members.
These messages are the ones used to reward the previous committee members. Intuitively, if a process receives a decision message for the decided block by a committee member, then probably that committee member followed the protocol during that height\footnote{That is not true in general, since Byzantine processes for instance can send the decided value at the end without doing anything during the execution.}.
This allows to implement the repeated consensus. 
In \cite{opodis18}, the authors formalized and proved correct the Tendermint repeated consensus algorithm, an example of committee-based blockchain protocol.

In this paper, we analyse the fairness of committee-based blockchains as described in the following section.

\section{Fairness of Committee-based Blockchains}\label{sec:fairness}

\paragraph{Chain quality and Committee-based Blockchains}
	In the blockchain literature, chain quality has been defined by \cite{GarayKL15}, and extended by \cite{PS17}, to study fairness in Bitcoin-like systems.
	A blockchain system has the property of chain quality if the proportion of blocks produced by honest processes in any given window, 
	is proportional to their relative mining power. 
	Intuitively, chain quality ensures that malicious processes do not produce more blocks than in expectation, given their mining power.
	One of the main differences between Bitcoin-like system and the committee-based blockchains is that in the first, one process produces a block, whereas in committee-based blockchain, a committee (a set) of processes produces a block. 
	A committee is not necessary composed only of correct processes, but can contain Byzantine or correct processes (with a correctness hypothesis of having some majority of members following the protocol).	
We cannot apply the definition of chain quality to committee-based blockchain.
Instead of defining the fairness with the blocks, we will define it relative to the proportion of total rewards a process gets.

Informally, we say that a blockchain protocol is fair if any \emph{correct process} (a process that followed the protocol) that has $\alpha$ fraction of the total merit in the system will get at least $\alpha$ fraction of the total reward that is given in the system. 
In order to tackle the fairness of a committee-based blockchain protocol such as HotStuff \cite{hotstuff19}, Hyperledger \cite{hyperledger}, Redbelly \cite{redbelly17}, SBFT \cite{sbft18} or Tendermint \cite{opodis18}, we split the mechanism in two: the \emph{selection mechanism} and the \emph{reward mechanism}.
We say that each process has a given merit, which represents the effort the process is putting to maintain the blockchain, for instance, it can represent the mining power of a process in proof-of-work blockchain.
The \emph{selection mechanism} selects for each new height the committee members (the processes that will run the consensus instance) for that height.
The \emph{reward mechanism}, is in charge to distribute rewards to committee members that produce a new block. 
Informally, if the selection mechanism is fair, then each process will become committee member proportionally to its merit; 
and if the reward mechanism is fair then for each height, only the correct committee members get a reward. 
By combining the two mechanisms, a correct process is rewarded at least a number proportional to its merit parameter over the infinite execution of the system. 
		\subsection{Selection Mechanism}\label{ssec:select_mech}
			\subsubsection{Definition and Fairness of Selection Mechanisms}

	In a system where the size of the committee is strictly lower than the size of the processes in the system, there should be a way to
	select the members of the committees.
	Always selecting the same processes is a way to centralize the system. 
	That set of processes can exercise a power of oligarchy, and add in the blockchain only transactions they want.
	
	Formally, a selection mechanism is the function $\selection: \Blocks^{+}\times \mathbb{N} \to \Pi^{n}\times \emptyset$, 
	where $n$ is the size of committees, and $\Blocks^{+}$ represents the set of non-empty blockchains such that 
	if $\bc$ is a blockchain, then:
		\[\selection(\bc,h) = \left\{\begin{array}{r r}
				V_h, 		&\text{if } |\bc|\ge h-1\\
				\emptyset,	&\text{otherwise}\\
			\end{array}\right. ,
		\]
		where we recall that $|\bc|$ means the length of the blockchain $|\bc|$, and $V_h$ is the set of committee members for height $h$.
		
	In the blockchain, some information can be saved, for instance the stakes of each process that represents their wealth.
	The selection mechanism can select, for instance, processes with the highest stakes, or processes with the lowest stakes,
	or always the same set of processes, or processes that were committee members less often, etc.
	
	To abstract the notion of effort of a process, we denote by $\alpha_i(t) \in [0,1]$ the merit parameter of $p_i$ at time $t$ 
	proportionally to the total merit at time $t$, such that $\forall p_i\in \Pi_\rho, \forall t, \alpha_i(t)$.
	
	If $\forall t \in \mathbb{N}, \forall i, \alpha_i(t) = \alpha_i(0)$, we denote by $\alpha_i$ the merit of the process $p_i$.
	That means that the merits do not depend on the evolution of the blockchain, nor on it contents\footnote{With the analogy of proof-of work blockchains,
	it means that the computing power of each processes do not change with time.}.
	Let $v_i$ be the number of times $p_i$ becomes committee member, proportionally to the number of blocks, so $v_i \in [0,1]$.
	We propose the following definition of fairness of selection mechanisms where merits are fixed and do not change over time. 
	The definition allows all processes with positive merit to be member of committees infinitely often, 
	and by doing so, respecting their merit.
	
	\begin{definition}[Fairness of Selection Mechanism]\label{def:fairness}
		Assume that the blockchain is built infinitely, so $\forall h\ge0$, there is a block at position $h$.
		We say that a selection mechanism is fair if it respects the following properties:
		\begin{enumerate}
			\item\label{selection1} If $\alpha_i \neq 0$ then $v_i \neq 0$; or equivalently, $\alpha_i \neq 0 \implies \forall h\ge 0, \exists h'\ge h: p_i \in V_{h'}$
			\item\label{selection2} If $\alpha_i \ge \alpha_j$ then $v_i \ge v_j$.
		\end{enumerate}
	\end{definition}
	Informally, Condition \ref{selection1} means that each process with a positive merit parameter should become a committee member infinitely often.
	Condition \ref{selection2} means that a process with a low merit cannot be selected more than a process with a higher merit.
	Note that this definition depends only on the merit and not on the behavior of the processes (correct or Byzantine).
	
	A definition of fairness of a generic selection mechanism (that does change over time) is still an open question, 
	but such definition should encapsulate the Definition \ref{def:fairness} as special case.
	
	\begin{remark}\label{r:selectAll}
		If the total number of processes in the system is equal to the the size of committees,
		then all processes are always selected, so the selection mechanism in that case is trivially fair,
		although asking a huge set of processes to run the consensus is not scalable.
	\end{remark}
			\subsubsection{Examples of Selection Mechanisms}	

In this section, we briefly study different possible selection mechanisms.
Let us assume that there are $N>n$ processes during the whole execution of the system.
Let us assume that merits do not change over time, $\forall t \in \mathbb{N}, \forall i, \alpha_i(t) = \alpha_i(0)$
and that all processes have the same merit, $\forall i, j \in \Pi_\rho, \alpha_j = \alpha_j>0$.

In this analysis, we consider selection mechanisms that depends on the current state of the  
stakes and/or number of selections of processes.
All processes are correct, and a committee member is rewarded once its committee produces a block.


For our analysis, we further consider that the processes are ordered by their stake and their id (public address for instance).
Let us assume that at the beginning of the execution, all processes have the same amount of stake. 
Without loss of generality, and up to renaming, consider also that during the execution, 
if $\exists i, j: i<j$ such that $p_i$ and $p_j$ have the same amount of
stakes, then $p_i$ is selected before $p_j$.


\paragraph{Select the processes with the highest stake}
	This selection mechanism works as follows:
	For any height $h$, with respect to the blockchain up to height $h-1$, 
	the $n$ processes having the biggest amount of stakes
	are selected to be part of the committee.
	This mechanism lead to the situation where only the $n$ processes selected first will always be selected,
	and the other processes will never be.
	This mechanism is not fair, since Condition \ref{selection1} is not satisfied.
	The processes not selected have by assumption a positive stake, and so they should be selected
	infinitely often.
	
	Note that in the special case where there are at most $n$ processes with a positive stake, 
	and these are all selected, then fairness is satisfied.

\paragraph{Select the processes with lowest stake}
	
	
	This selection mechanism works as follows:
	for any height $h$, with respect to the blockchain up to height $h-1$, 
	the $n$ processes having the lowest amount of stakes
	are selected to be part of the committee.
	If all $N$ processes are part of the system since the beginning, 
	the number of
	times each process has been selected after $l$ blocks, 
	on average is 
	$l*n/N$ selections.
	This mechanism is fair according to the Definition \ref{def:fairness}.
	
\begin{remark}\label{r:importanceFairnessSelection}
	An unfair selection mechanism can lead to a centralization of the system, by always letting the same processes decide on the next block.
	Although the assumption on the bound of Byzantine processes does not depend on the selection mechanism, we note that
	when a selection mechanism selected the processes with the lowest amount of stakes, and only correct processes in committees are rewarded,
	at some point processes that were non-correct will have the least stake, and thus be selected. 
\end{remark}


		\subsection{Reward Mechanism}\label{ssec:rewmech}

	\subsubsection{Definition of Reward Mechanism}
		In blockchain systems, a process that produces and adds blocks to the blockchain are rewarded.
		In a committee-based blockchain, a committee is the producer of the block.
		Within that committee, some processes may not behave as prescribed.
		There may be different ways of rewarding members of committees.
		To do so, we define the \emph{reward mechanism}. A reward mechanism consists of the reward function defined as follows:
		$\reward: \Blocks^{+}\times \mathcal{P}(\textit{Information}) \times \mathbb{N} \to \mathbb{R}^{|\Pi|}\times (\bot, \dots, \bot)$,
		where $\mathcal{P}(\textit{Information})$ is the power set of all messages.
		
		If $\bc$ is a blockchain and $|\bc|< h$, $\reward(\bc,\info,h) = (\bot, \dots, \bot)$.
		Otherwise, it assigns to each process a given reward, null or not.
		In $\reward(\bc,\info,h)$, $\info$ represents the set of messages received, $h$ the height of the blockchain $\bc$ 
		where the reward of committee member is computed.
		
		A reward is considered allocated if it is written in the blockchain. 
		The second part of the reward mechanism is to choose when to
		allocate the rewards corresponding for a given height.
		If a reward has been allocated at a height $h$, the process can use it after a certain number of blocks defined in the genesis block (e.g. \cite{GarayKL15,ES18}).
		We consider that for each production of block its rewards are allocated in one block and not over different blocks, 
		such that after the allocation of rewards
		a process knows if it has been rewarded or not.
		Note that once rewards are allocated, they cannot change any more.
		Some blockchains system adds a system of punishment, called \emph{slashing}, 
		to afflict, afterwards, some costs to a process if there is a proof of some misbehavior,
		as described in \cite{OG19}.

	\subsubsection{Fairness of Reward Mechanism}
		We define the following properties for characterizing the fairness of a reward mechanism. 
		Let $h$ be a height. 
		Recall that $p_i$ is considered $h$-correct if during the period of height $h$, $p_i$ followed the protocol.
		Each committee member has a boolean variable $r_i^h$, which we call \emph{reward parameter} defined as follows:
		\begin{enumerate}
			\item \label{fairness1} If $p_i$ is a not a committee members for $h$, then $r_i^h=0$,
			\item \label{fairness2} \textbf{$h$-completeness.} If $p_i$ is a committee members for $h$ and $p_i$ is $h$-correct, then $r_i^h=1$,
			\item \label{fairness3} \textbf{$h$-accuracy.} If $p_i$ is a committee members for $h$ and $p_i$ is not $h$-correct, then $r_i^h=0$.
			


		\end{enumerate}
		If $r_i^h=0$, it means that $p_i$ is not rewarded for height $h$, and if $r_i^h=1$, $p_i$ has been rewarded for $h$.
		The properties are inspired by classical properties of failure detectors \cite{CT96}.
		
		\begin{remark}
			Note that we do not reward non-committee members. 
			In this article, we do not consider delegations. When a process delegates to a committee member, 
			once the committee member is rewarded, all of its delegates are rewarded proportionally to what they delegated.
			In future works, we may consider the case of committee-based blockchains with delegations. 
			To do so, $r_i^h$ must contain more information and not just be a boolean variable.
		\end{remark}
		
		\begin{definition}[Complete Fairness of a Reward Mechanism]
			Let $\mathcal{R}$ be a reward mechanism. 
			If $\forall h>0$, $\mathcal{R}$ satisfies conditions \ref{fairness1} and $h$-completeness (condition \ref{fairness2}),
			we say that $\mathcal{R}$ satisfies complete fairness.
		\end{definition}
		If a reward mechanism satisfies complete fairness, it means that for all height $h>0$, 
		all $h$-correct committee members are rewarded, and non-committee members are not. 
		
		\begin{proposition}\label{prop:completeFairness}
			There always exists at least one reward mechanism satisfying complete fairness. 
		\end{proposition}
		Once a block is in the chain, 
		rewarding all committee members, in the next block, for that block and only them,  satisfy conditions \ref{fairness1} and \ref{fairness2}.
		Condition \ref{fairness1} is satisfied since for all height, non-committee members are not rewarded. 
		Condition \ref{fairness2} also holds, for any given height $h$, all committee members of $h$ are rewarded, 
		in particular all $h$-correct committee members.

		\begin{definition}[Accurate Fairness of a Reward Mechanism]
			Let $\mathcal{R}$ be a reward mechanism. 
			If $\forall h>0$, $\mathcal{R}$ satisfies conditions \ref{fairness1} and $h$-accuracy (condition \ref{fairness3}),
			we say that $\mathcal{R}$ satisfies accurate fairness.
		\end{definition}
		If a reward mechanism satisfies accurate fairness, it means that for all height $h>0$, all non $h$-correct committee members are not rewarded. 
		
		\begin{proposition}\label{prop:accurateFairness}
		There always exists at least one reward mechanism satisfying accurate fairness. 
		\end{proposition}
		Never allocating rewards satisfies conditions \ref{fairness1} and \ref{fairness3}.
		Condition \ref{fairness1} is satisfied since non-committee members are not rewarded.
		Condition \ref{fairness3} holds, since no process is rewarded. 
		In particular for any given height $h>0$, 
		no non $h$-correct committee members is rewarded.
		
		Although simple and trivial to satisfy either complete fairness or accurate fairness, 
		satisfying both at the same time is more complex and not always possible.
		
		\begin{definition}[Fairness of a Reward Mechanism]
			Let $\mathcal{R}$ be a reward mechanism. 
			If $\forall h>0$, $\mathcal{R}$ satisfies conditions \ref{fairness1}, $h$-completeness (condition \ref{fairness2}) and $h$-accuracy (condition \ref{fairness3}),
			we say that $\mathcal{R}$ is fair.
		\end{definition}
		
		We say that a reward mechanism is fair when at each height $h$, all and only $h$-correct committee members are rewarded.

		\begin{remark}
			If $\forall h>0, |V_h| > 1$ then for a reward mechanism to be (eventually) fair, rewards cannot be allocated directly in the block. For any height $h> 0$ the set of $h$-correct committee members cannot be known in advance.
			If $\forall h>0, |V_h| = 1$, the reward can be directly given to the only committee member, so in the block at height $h$.
		\end{remark}
		
		\begin{theorem}\label{t:strongFairness}
			There exists a fair reward mechanism in a committee-based blockchain protocol iff the system is synchronous.
		\end{theorem}
		\begin{proofT}
			We prove this theorem by double implication.
			\begin{itemize}
				\item If there exists a fair reward mechanism, then the system is synchronous.
				
					Let $\mathcal{R}$ be a reward mechanism. By contradiction, we assume that $\mathcal{R}$ is fair and that the system is not synchronous. 
					
					$V_{h}$ is the set of committee members for the height $h$. Let $k>0$ be the number of blocks to wait before distributing the rewards for $V_h$. The reward is allocated by the committee $V_{h'}$, where $h'=h+k$.
					Since the system is not synchronous, the committee members of height $h'$, $V_{h'}$, may not receive all messages from $V_{h}$ before allocating the rewards.
					
					By conditions \ref{fairness1}, and \ref{fairness2}, 
					since the reward mechanism is fair, by conditions \ref{fairness1} - \ref{fairness3},
					all and only the $h$-correct committee members of the height $h$ have a reward parameter equal to $1$.
					That means that the $h'$-correct committee members of $V_{h'}$ know exactly 
					who were the $h$-correct committee members in $V_h$, so
					they got all the messages before giving the reward. Contradiction, so the system is synchronous.
					
				\item If the system is synchronous, then there exists a fair reward mechanism.
					
					We assume that the system is synchronous and $\forall h>1$, all messages sent by $h$-correct processes at height $h$ are delivered before the block at height $h+1$. 
					Let $\mathcal{R}$ be the following reward mechanism:
					let $h$ be a height. Rewards for a block at height $h$ are allocated at height $h+1$  by the committee $V_{h+1}$.
					\begin{itemize}
						\item If a process is not a committee member for height $h$, 
							set its reward parameter to $0$, this is known since processes are already at height $h+1$.
						\item By combining the messages from committee member of $h$ processes, since the communication system is synchronous, it is possible to differentiate between $h$-correct and non $h$-correct committee members, and so set the reward parameter of $h$-correct committee members of $h$ to $1$; and set the reward parameter of non $h$-correct committee members of $h$ to $0$.
					\end{itemize}
%
					By construction, the committee members in $h+1$ allocates rewards to all and only $h$-correct committee members, so $\mathcal{R}$ is fair, because it satisfies all fairness conditions \ref{fairness1} - \ref{fairness3}.
			\end{itemize}
			\renewcommand{\toto}{t:strongFairness}
		\end{proofT}
		
		If there is no synchrony, then there cannot be a fair reward mechanism for committee-based blockchains. 
		Our definition of fairness states that for any height conditions \ref{fairness1}-\ref{fairness3} are satisfied. 
		At any time, processes receive all rewards they deserve.
		This definition can be weakened.
		
		\begin{definition}[Eventual Fairness of a Reward Mechanism]
			Let $\mathcal{R}$ be a reward mechanism. 
			If $\exists h_0>0: \forall h \ge h_0$, $\mathcal{R}$ satisfies conditions \ref{fairness1},  $h$-completeness (condition \ref{fairness2}) and $h$-accuracy (condition \ref{fairness3}),
			we say that $\mathcal{R}$ is eventually fair.
		\end{definition}
		
		A reward mechanism is eventually fair if after an unknown time, the rewards are allocated to and only 
		to correct committee members.
		
		We note that if a reward mechanism is fair, then it is eventually fair but the reverse (reciprocal) is not necessarily true.

		\paragraph*{Detectable Byzantine Processes}
			
			In synchronous systems,
			it is always possible to detect Byzantine processes, for example
			using the broadcast abstraction detectable all-to-all (DA2A)
			defined in \cite{LSKM19}. Detecting Byzantine processes allows to not
			reward them, and then to satisfy condition \ref{fairness3}. 
			On the other hand, in eventual
			synchronous systems, the problem is more
			difficult. Kihlstrom et al., in \cite{KMM03} distinguish
			between detectable and non-detectable Byzantine. The detectable
			Byzantine are the processes whose behavior can
			be detected, for instance by doing omission or commissions
			failures. Non-detectable Byzantine, are Byzantine processes
			whose fault cannot be detected, for example the processes that
			alter their internal state. In \cite{GLAS12}, Greve et al. further extend that
			approach to propose a failure detector for detectable Byzantine
			in dynamic networks. In line with \cite{KMM03}, we say that a Byzantine
			is \emph{detectable} if it is possible for correct processes to detect it
			by combining information they have (e.g what they
			see during the execution).
			
			
			Note that although Kihlstrom \emph{et al.} proposed in \cite{KMM03} a failure detector for solving consensus, 
			our problem is not the same, and we cannot apply their failure detector as it is.
			In \cite{KMM03}, once a process has a detectable Byzantine behavior, it should be suspected forever.
			In our model, we do not want to punish indefinitely Byzantine processes. 
			In fact, we want for any height $h$ to not reward only processes that were not $h$-correct.
			For example, let $p_i$  be a process such that it is part of committees $h$ and $h'$, such that $h' > h$.
			Suppose also that during height $h$, $p_i$ sends contradictory messages (and so is Byzantine), but follows the
			protocol during height $h'$. If $p_i$ has been detected and not rewarded for height $h$, that should be taken into consideration
			of its work during height $h'$, and since it follows the protocol, it should be rewarded for height $h'$.
			The failure detector proposed by Kihlstrom \emph{et al.}, once a process has been suspected, marks such process as Byzantine forever.
		
		\begin{theorem}\label{t:eventualFairness}
			There exists an eventual fair reward mechanism in a committee-based blockchain protocol iff the system is (eventually) synchronous and Byzantine processes are detectable.
		\end{theorem}
		\begin{proofT}
			We prove this theorem by double implication.
			\begin{itemize}
				\item If there exists an eventual fair reward mechanism, then the system is eventually synchronous or synchronous and Byzantine processes are detectable.
				
					Let $\mathcal{R}$ be a reward mechanism. We assume that $\mathcal{R}$ is eventually fair.

					If $\mathcal{R}$ is fair, by Theorem \ref{t:strongFairness}, the communication is synchronous, which ends the proof.
					Otherwise, since $\mathcal{R}$ is eventually fair, that means that there is a point in time $h$ from which all the rewards are correctly allocated, so for any height $h'\ge h$, $h'$-correct committee members of committees at height $h'$ are able to distinguish between non-correct processes during the height they are distributing the rewards, the Byzantine are then detectable.
					If we consider $h$ as the beginning of the execution, then we have that $\mathcal{R}$ is fair, 
					and by Theorem \ref{t:strongFairness}, the message delay is upper bounded.
					We have that after $h$, the message delay is upper bounded, so the communication is eventually synchronous.
					So the Byzantine are then detectable, and the communication is synchronous or eventually synchronous.
					
				\item If the system is eventually synchronous or synchronous, and Byzantine processes are detectable, then there exists an eventual fair reward mechanism.
					
					If the system is synchronous, the proof follows directly from Theorem \ref{t:strongFairness}.
					Consider that the system is eventually synchronous, but not synchronous.
					Let $\mathcal{R}$ be the following mechanism:
					Let $h$ be a height. Rewards for a block at height $h$ are allocated at height $h+1$ by the committee $V_{h+1}$.
					\begin{itemize}
						\item If a process is not a committee member for height $h$, 
							set its reward parameter to $0$, this is known since processes are already at height $h+1$.
						\item By combining the messages from committee member of $h$ processes, if there is not sufficient information to detect the behavior of processes, reward only those detected as $h$-correct and in $V_h$, and the process proposing the distribution of reward increases the duration to wait before starting the next height ($\Delta$ in $\Figure$ \ref{fig:repConsensus}) .
						If there is enough information to detect the behavior of all processes in $V_h$, then reward the $h$-correct processes in $V_h$ and do not reward non $h$-correct processes in $V_h$.
					\end{itemize}
					$\mathcal{R}$ is eventually fair.
%
			\end{itemize}
			\renewcommand{\toto}{t:eventualFairness}
		\end{proofT}

	 	\begin{corollary}\label{c:asyncFairness}
			In an asynchronous system, there is no (eventual) fair reward mechanism in a committee-based blockchain tolerating Byzantine processes.
		\end{corollary} 
		\begin{proofC}
			Assume that the system is asynchronous, where there are good periods such that consensus can be reached.
			By contradiction, let $\mathcal{R}$ be an eventual fair reward mechanism.
			\begin{itemize}
				\item If there are non-detectable Byzantine processes in the system, then by Theorem \ref{t:eventualFairness}, $\mathcal{R}$ is not fair;
				\item If all Byzantine processes are detectable, then by Theorem \ref{t:eventualFairness}, the system must be synchronous, or eventually synchronous.
			\end{itemize}
			We have a contradiction, since the system is asynchronous. 
			It is not possible to have an (eventual) fair reward mechanism in an asynchronous system.
			
			\renewcommand{\toto}{c:asyncFairness}
		\end{proofC}
			Note that this result holds even if all the processes are correct, but not known in advance, and the protocol tolerates Byzantine faults.
			In fact, it is different of the FLP impossibility result of consensus in an asynchronous system with one faulty process \cite{FLP}.


\section{Numerical Examples}\label{sec:simulation}

In this section, we examine the impact of different communication
models on the fairness of reward mechanisms through
several numerical examples that confirm the results on
the fairness of reward mechanisms from section \ref{ssec:rewmech}.

\paragraph{Execution}
	In our analyses, processes run a committee-based blockchain protocol 
	as described in section \ref{sec:algo}, and rewards for a block produced at a height $h$ are allocated in the block at height $h+1$.
	$\Figure$ \ref{fig:repConsensus} depicts the state machine of the execution.
	Note that the consensus module is Byzantine fault tolerant.
	We highlight the environment's important characteristics:  the communication system, the total number of processes in the system,
	the size of each committee,
	the different type of processes and their number at a given height,
	the rewarding mechanism, and 
	the selection mechanism.
	We must choose the value of these parameters before launching the execution.
	We consider different communication systems, and rewards are allocated by the next committee by using
	messages they delivered from the previous height -- 
	use the combination of all messages and check if they correspond to the correct time and a possible value to send according to the state.

We consider a system where all processes are part of all consensus instances.
For clarity, and without loss of generality, we consider a system with $n\ge 4$ processes where all are selected. 
There is no impact from the selection mechanism. 
As stated in Remark \ref{r:selectAll}, selecting all processes is a fair selection mechanism.
We can now focus on the impact of the network on rewards.
For any height $h$, there can be at most $\lfloor (n-1)/3 \rfloor$ non $h$-correct processes in each committee.
For a committee, a quorum of $\lceil 2n/3\rceil$ is needed for any decisions.
In the case where there are for any height $h$ some non $h$-correct processes, we assume that processes have
enough information to detect it when distributing rewards. In particular, and for the experiment the Byzantine are specially tagged,
and that tag is used only for allocating rewards. 
When an $h$-correct receives a message from a non $h$-correct, it suspects it, and broadcasts the information.
When an $h$-correct delivers at least $2\lfloor n/3\rfloor+1$ suspicions for a process, it considers it as non $h$-correct, and does not propose to reward it.

We use MATLAB \cite{matlab} for our analysis.
We consider three different communication models.
First, we consider a synchronous communication, where there is no delay.
Then we consider the two following semi-synchronous communication models 
(i) the system alternate between sufficiently good and bad periods where during good periods, 
messages delays are upper bounded (good/bad model), and
(ii) from an unknown time message delays are upper bounded (eventually synchronous model).

In all these models, consensus can be reached.
In the good/bad model, progress for consensus instances are guaranteed during the good periods.
Note that in the eventually synchronous model, once the global stabilization time (GST) happens
all message delays are upper bounded. If for a process, GST happens during height $h$, then for all height $h'>h$,
the message delays are upper bounded.

In each configuration of the communication model, we ran the experiment 50 times, and took the mean.
We represent by $0$ if a process did not receive a reward, and $1$ if the process received a reward for the  corresponding height.

\subsection{Synchronous Model}
	In the synchronous system, we consider that processes receive messages instantly; there is no message delay.
	
	When all processes are correct in a synchronous system, all processes are always rewarded.
	Since messages are received instantly, all processes always have the same information.
	If for a height $h$, one process is non $h$-correct, other processes have sufficient information to detect it,
	the non $h$-correct is not rewarded, and it is true for all height.
	Since for all height $h>0$, all $h$-correct processes have a reward, and non $h$-correct processes do not have a reward, 
	the reward mechanism is fair. The conditions \ref{fairness1}, $h$-completeness (condition \ref{fairness2}) and $h$-accuracy (condition \ref{fairness3}) of reward mechanisms are satisfied.
	
\subsection{Good/Bad Periods Model}	
	
	We consider a system where there are good and bad periods, 
	where good periods are sufficiently long in order to achieve consensus.
	We also assume that in this setting all processes are correct throughout the whole execution.
	If the reward mechanism is (eventually) fair, then all processes should always be rewarded (from a time on), since they are all correct.
	That is not the case. 
	For instance, consider the following scenario:
	the system is composed of $4$ processes, all correct. 
	Moreover, all processes but one, $p_4$, are well connected and such that consensus is always reached.
	$p_4$ eventually reaches consensus, but always after the other processes decide on the set of processes to reward.
	$p_4$ will never be rewarded, although it always participates correctly to each consensus instance.
	The $h$-completeness condition \ref{fairness2} is never satisfied because of $p_4$, so the mechanism is neither fair, nor eventually fair.
	
\subsection{Eventually Synchronous Model}
	\begin{figure}[t!]
		\centering
			\includegraphics[width=\linewidth]{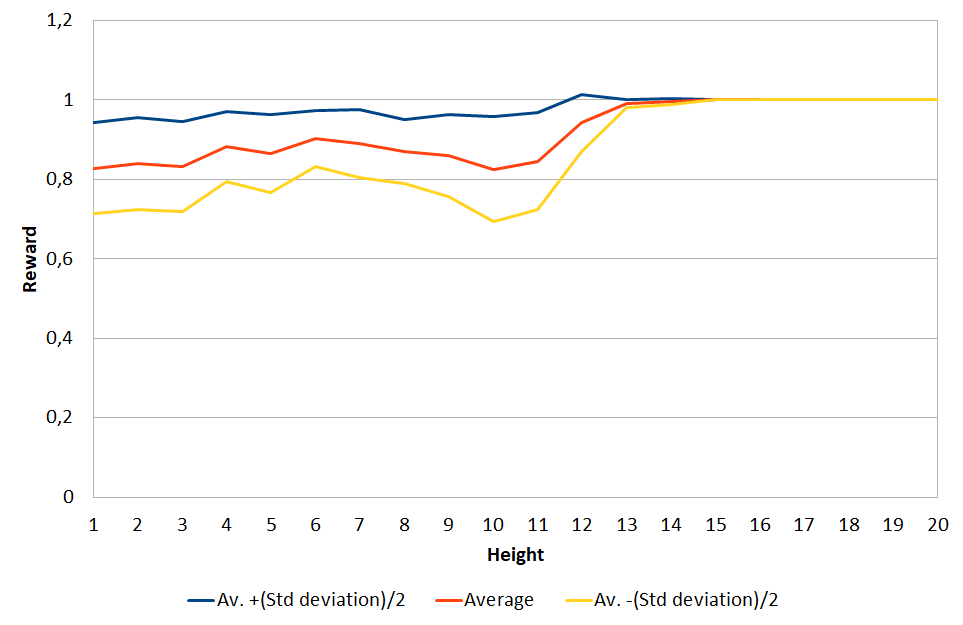}
		\caption{Evolution of Rewards in an Eventual Synchronous System, where Global Stabilization Time happens during height $10$.}
		\label{fig:simuEvSynchronous}
	\end{figure}
	We consider an eventual synchronous model where all processes are correct, and part of each committee instances. 
	Recall that eventual synchronous is a system where after a finite but unknown time (the Global Stabilization Time), message delay is upper bounded for the rest of the execution.
	In our examples, we consider that the Global Stabilization Time happens during height $10$.
	
	On $\Figure$ \ref{fig:simuEvSynchronous}, we show the evolution of reward  for each height.
	We draw the mean of the average reward of each participant (represented by the red curve).
	The blue and yellow curves represent the standard deviation. We can see the set in which the processes are rewarded.
	If the blue and yellow curves converge, it means that all processes have on average the same reward.
	
	On $\Figure$ \ref{fig:simuEvSynchronous}, 
	from height $10$, the evolution of the reward is increasing.
	Approximately from height $14$ till the end, all processes are rewarded.
	Before height $10$, there is a fluctuation in the evolution of rewards allocated because of the asynchronous periods, 
	processes are not necessarily rewarded even if the participate. Their messages were not received on time.
	Once the global stabilization time happens, the message delays become upper bounded, 
	but some processes still have a time-out shorter than the bound. 
	These processes still increase their time-out until they receive all messages, or detect incorrect behavior.
	When all processes deliver messages during their corresponding rounds, they allocate the rewards to all correct processes.
	All processes being correct, it means that the reward mechanism is eventually fair, which happens in our example.

\section{Conclusions and Future Works}\label{sec:conclusion}


The originality of our contribution  is the study of the impact of
network conditions on the fairness of the rewarding in committee-based
blockchains prone to Byzantine behavior.
We proved that the reward mechanism  is (eventually) fair iff the system
communication is (eventually) synchronous and Byzantine
processes are detectable. Our study opens interesting future
research directions in particular the extension to other types
of behaviors such as rational or amnesic. Furthermore, we are interested
in studying
the impact of network attacks on the fairness of rewarding.
Another interesting direction is the design of self-adaptative fair
rewarding schemes.
	
\section*{Acknowledgment}
The authors thank Ludovic Desmeuzes for his work on the numerical examples.


\bibliographystyle{plain}
\bibliography{biblio}

\begin{thebibliography}{10}

\bibitem{aguilera2004}
Marcos~K Aguilera.
\newblock A pleasant stroll through the land of infinitely many creatures.
\newblock {\em ACM Sigact News}, 35(2):36--59, 2004.

\bibitem{opodis18}
Yackolley Amoussou{-}Guenou, Antonella {Del Pozzo}, Maria Potop{-}Butucaru, and
  Sara Tucci{-}Piergiovanni.
\newblock {Correctness of Tendermint-Core Blockchains}.
\newblock In {\em {OPODIS} 2018, December 17-19, 2018, Hong Kong, China}, pages
  16:1--16:16, 2018.

\bibitem{netys19}
Yackolley Amoussou{-}Guenou, Antonella~Del Pozzo, Maria Potop{-}Butucaru, and
  Sara {Tucci Piergiovanni}.
\newblock Dissecting tendermint.
\newblock In {\em Networked Systems - 7th International Conference, {NETYS}
  2019, Marrakech, Morocco, June 19-21, 2019, Revised Selected Papers}, pages
  166--182, 2019.

\bibitem{ADLPT19}
Emmanuelle Anceaume, Antonella {Del Pozzo}, Romaric Ludinard, Maria
  Potop{-}Butucaru, and Sara Tucci{-}Piergiovanni.
\newblock {Blockchain Abstract Data Type}.
\newblock In {\em The 31st {ACM} on Symposium on Parallelism in Algorithms and
  Architectures, {SPAA} 2019, Phoenix, AZ, USA, June 22-24, 2019.}, pages
  349--358, 2019.

\bibitem{hyperledger}
Elli Androulaki, Artem Barger, Vita Bortnikov, Christian Cachin, Konstantinos
  Christidis, Angelo~De Caro, David Enyeart, Christopher Ferris, Gennady
  Laventman, Yacov Manevich, Srinivasan Muralidharan, Chet Murthy, Binh Nguyen,
  Manish Sethi, Gari Singh, Keith Smith, Alessandro Sorniotti, Chrysoula
  Stathakopoulou, Marko Vukolic, Sharon~Weed Cocco, and Jason Yellick.
\newblock {Hyperledger Fabric: A Distributed Operating System for Permissioned
  Blockchains}.
\newblock In {\em Proceedings of the Thirteenth EuroSys Conference, EuroSys
  2018, Porto, Portugal, April 23-26, 2018}, pages 30:1--30:15, 2018.

\bibitem{CT96}
Tushar~Deepak Chandra and Sam Toueg.
\newblock Unreliable failure detectors for reliable distributed systems.
\newblock {\em J. {ACM}}, 43(2):225--267, 1996.

\bibitem{redbelly17}
Tyler Crain, Vincent Gramoli, Mikel Larrea, and Michel Raynal.
\newblock {(Leader / Randomization / Signature)-free Byzantine Consensus for
  Consortium Blockchains}, 2017.

\bibitem{PeerCensus}
Christian Decker, Jochen Seidel, and Roger Wattenhofer.
\newblock Bitcoin meets strong consistency.
\newblock In {\em Proceedings of the 17th International Conference on
  Distributed Computing and Networking, Singapore, January 4-7, 2016}, pages
  13:1--13:10, 2016.

\bibitem{DDFPT08}
Carole Delporte{-}Gallet, St{\'{e}}phane Devismes, Hugues Fauconnier, Franck
  Petit, and Sam Toueg.
\newblock With finite memory consensus is easier than reliable broadcast.
\newblock In {\em Principles of Distributed Systems, 12th International
  Conference, {OPODIS} 2008, Luxor, Egypt, December 15-18, 2008. Proceedings},
  pages 41--57, 2008.

\bibitem{DLS88}
Cynthia Dwork, Nancy~A. Lynch, and Larry~J. Stockmeyer.
\newblock Consensus in the presence of partial synchrony.
\newblock {\em J. {ACM}}, 35(2):288--323, 1988.

\bibitem{ES14}
Ittay Eyal and Emin~G{\"{u}}n Sirer.
\newblock Majority is not enough: Bitcoin mining is vulnerable.
\newblock In {\em Financial Cryptography and Data Security - 18th International
  Conference, {FC} 2014, Christ Church, Barbados, March 3-7, 2014, Revised
  Selected Papers}, pages 436--454, 2014.

\bibitem{ES18}
Ittay Eyal and Emin~G{\"{u}}n Sirer.
\newblock Majority is not enough: bitcoin mining is vulnerable.
\newblock {\em Commun. {ACM}}, 61(7):95--102, 2018.

\bibitem{FKORVW18}
Giulia~C. Fanti, Leonid Kogan, Sewoong Oh, Kathleen Ruan, Pramod Viswanath, and
  Gerui Wang.
\newblock Compounding of wealth in proof-of-stake cryptocurrencies.
\newblock {\em CoRR}, abs/1809.07468, 2018.

\bibitem{FLP}
M.~J. Fischer, N.~A. Lynch, and M.~S. Paterson.
\newblock Impossibility of distributed consensus with one faulty process.
\newblock {\em Journal of the ACM}, 32(2), April 1985.

\bibitem{francez86}
Nissim Francez.
\newblock {\em Fairness}.
\newblock Texts and Monographs in Computer Science. Springer, 1986.

\bibitem{GarayKL15}
J.~A. Garay, A.~Kiayias, and N.~Leonardos.
\newblock The bitcoin backbone protocol: Analysis and applications.
\newblock In {\em Proc. of the EUROCRYPT International Conference}, 2015.

\bibitem{sbft18}
Guy Golan{-}Gueta, Ittai Abraham, Shelly Grossman, Dahlia Malkhi, Benny Pinkas,
  Michael~K. Reiter, Dragos{-}Adrian Seredinschi, Orr Tamir, and Alin Tomescu.
\newblock {SBFT:} a scalable decentralized trust infrastructure for
  blockchains.
\newblock {\em CoRR}, abs/1804.01626, 2018.

\bibitem{GLAS12}
Fab{\'{\i}}ola Greve, Murilo~Santos de~Lima, Luciana Arantes, and Pierre Sens.
\newblock A time-free byzantine failure detector for dynamic networks.
\newblock In {\em 2012 Ninth European Dependable Computing Conference, Sibiu,
  Romania, May 8-11, 2012}, pages 191--202, 2012.

\bibitem{GW18}
Rachid Guerraoui and Jingjing Wang.
\newblock On the unfairness of blockchain.
\newblock In {\em Networked Systems - 6th International Conference, {NETYS}
  2018, Essaouira, Morocco, May 9-11, 2018, Revised Selected Papers}, pages
  36--50, 2018.

\bibitem{GRT18}
{\"{O}}nder G{\"{u}}rcan, Alejandro~Ranchal Pedrosa, and Sara {Tucci
  Piergiovanni}.
\newblock On cancellation of transactions in bitcoin-like blockchains.
\newblock In {\em On the Move to Meaningful Internet Systems. {OTM} 2018
  Conferences - Confederated International Conferences: CoopIS, C{\&}TC, and
  {ODBASE} 2018, Valletta, Malta, October 22-26, 2018, Proceedings, Part {I}},
  pages 516--533, 2018.

\bibitem{GDT17}
{\"{O}}nder G{\"{u}}rcan, Antonella~Del Pozzo, and Sara {Tucci Piergiovanni}.
\newblock On the bitcoin limitations to deliver fairness to users.
\newblock In {\em On the Move to Meaningful Internet Systems. {OTM} 2017
  Conferences - Confederated International Conferences: CoopIS, C{\&}TC, and
  {ODBASE} 2017, Rhodes, Greece, October 23-27, 2017, Proceedings, Part {I}},
  pages 589--606, 2017.

\bibitem{HM16}
Maurice Herlihy and Mark Moir.
\newblock Enhancing accountability and trust in distributed ledgers.
\newblock {\em CoRR}, abs/1606.07490, 2016.

\bibitem{KKNZ19}
Dimitris Karakostas, Aggelos Kiayias, Christos Nasikas, and Dionysis Zindros.
\newblock {Cryptocurrency Egalitarianism: A Quantitative Approach}.
\newblock In {\em The 1st International Conference on Blockchain Economics,
  Security and Protocols, {Tokenomics} 2019, Paris, France, May 06-07, 2019},
  2019.

\bibitem{KRDO17}
Aggelos Kiayias, Alexander Russell, Bernardo David, and Roman Oliynykov.
\newblock Ouroboros: {A} provably secure proof-of-stake blockchain protocol.
\newblock In {\em Advances in Cryptology - {CRYPTO} 2017 - 37th Annual
  International Cryptology Conference, Santa Barbara, CA, USA, August 20-24,
  2017, Proceedings, Part {I}}, pages 357--388, 2017.

\bibitem{KMM03}
Kim~Potter Kihlstrom, Louise~E. Moser, and P.~M. Melliar{-}Smith.
\newblock Byzantine fault detectors for solving consensus.
\newblock {\em Comput. J.}, 46(1):16--35, 2003.

\bibitem{BizCoin}
E.~Kokoris{-}Kogias, P.~Jovanovic, N.~Gailly, I.~Khoffi, L.~Gasser, and
  B.~Ford.
\newblock {Enhancing Bitcoin Security and Performance with Strong Consistency
  via Collective Signing}.
\newblock In {\em Proceedings of the 25th {USENIX} Security Symposium}, 2016.

\bibitem{LSKM19}
Kfir Lev{-}Ari, Alexander Spiegelman, Idit Keidar, and Dahlia Malkhi.
\newblock Fairledger: {A} fair blockchain protocol for financial institutions.
\newblock {\em CoRR}, abs/1906.03819, 2019.

\bibitem{matlab}
MATLAB.
\newblock {\em version 9.6 (R2019a)}.
\newblock The MathWorks Inc., Natick, Massachusetts, 2019.

\bibitem{bitcoinNakamoto}
S.~Nakamoto.
\newblock {Bitcoin: A Peer-to-Peer Electronic Cash System}.
\newblock \url{https://bitcoin.org/bitcoin.pdf} (visited on 2019-08-15), 2008.

\bibitem{OG19}
Dev Ojha and Christopher Goes.
\newblock {F1 Fee Distribution}.
\newblock In {\em The 1st International Conference on Blockchain Economics,
  Security and Protocols, {Tokenomics} 2019, Paris, France, May 06-07, 2019},
  2019.

\bibitem{PS17}
Rafael Pass and Elaine Shi.
\newblock The sleepy model of consensus.
\newblock In {\em Advances in Cryptology - {ASIACRYPT} 2017 - 23rd
  International Conference on the Theory and Applications of Cryptology and
  Information Security, Hong Kong, China, December 3-7, 2017, Proceedings, Part
  {II}}, pages 380--409, 2017.

\bibitem{RA80}
M.~Pease, R.~Shostak, and L.~Lamport.
\newblock Reaching agreement in the presence of faults.
\newblock {\em Journal of the ACM}, 27(2):228--234, April 1980.

\bibitem{Saleh19}
Fahad Saleh.
\newblock Blockchain {{Without Waste}}: {{Proof}}-of-{{Stake}}.
\newblock {{SSRN Scholarly Paper}} ID 3183935, {Social Science Research
  Network}, {Rochester, NY}, January 2019.

\bibitem{hotstuff19}
Maofan Yin, Dahlia Malkhi, Michael~K. Reiter, Guy Golan{-}Gueta, and Ittai
  Abraham.
\newblock Hotstuff: {BFT} consensus with linearity and responsiveness.
\newblock In {\em Proceedings of the 2019 {ACM} Symposium on Principles of
  Distributed Computing, {PODC} 2019, Toronto, ON, Canada, July 29 - August 2,
  2019.}, pages 347--356, 2019.

\end{thebibliography}
\label{sec:biblio}

\newpage
%
\end{document}